\documentclass[twocolumn,pra,showpacs,preprintnumbers,amsmath,amssymb]{revtex4}

\usepackage{graphicx}
\usepackage{dcolumn}
\usepackage{bm}

\newcommand{\omperp}{{\omega}}
\newcommand{\gradphi}{\nabla{\phi_{j}}}

\begin{document}

\title{Dynamics of two-component Bose-Einstein condensates in rotating traps}

\author{I. Corro$^1$, R.G. Scott$^2$ and A.M. Martin$^1$}
\affiliation{
$^1$ School of Physics, University of Melbourne, Parkville, Victoria 3010, Australia. \\
$^2$ School of Physics and Astronomy, University of Nottingham, Nottingham NG7 2RD, U.K.}

\begin{abstract}
The dynamics of two-component Bose-Einstein condensates in rotating traps is investigated.  In the Thomas-Fermi limit, equations of motion are derived showing multiple static solutions for a vortex free condensate.  Dynamic stability analysis of these solutions and comparison with Truncated Wigner simulations enables us to identify the regimes for which vortex states will occur. In addition, our analysis predicts centre-of-mass oscillations that are induced by interspecies interactions and affect each component separately.  For attractive interspecies interactions, these oscillations lead to a stable symmetry broken state.
\end{abstract}

\maketitle
A two-component Bose-Einstein Condensate (TCBEC) exhibits a wide range of interesting behaviour that has been the subject of much theoretical and experimental research.  The two components may form either miscible or immiscible phases exhibiting complex density profiles \cite{Ho-PRL-1996,Riboli-PRA-2002,Pu_Bigelow-PRL-1998,Esry_PRL-1997, Kasamatsu_Yatsui_Tsubota-PRA2001}. For repulsive interspecies interactions, the two components may also form symmetry broken states \cite{Esry_PRL-1997,Hall_PRL1998,Esry-Greene_PRA1999,Ohberg-Stenholm_PRA1998, Kasamatsu_Yatsui_Tsubota-PRA2001, Gordon-savage_PRA1998,Chui-Ao_PRA1999}.
For rotating TCBECs, interlocking vortex lattices \cite{Mueller_Ho-PRL-2002} and vortex sheets \cite{Kasamatsu_PRL2003} have been predicted theoretically, with vortex lattices being confirmed experimentally \cite{Schweikhard-PRL-2004}.  This has led to much interest in the dynamics, and collective excitations of these systems \cite{kasamatsu-2005-19,Woo_PRAR2007,Barnt_NJP2008}, and recently a number of papers have also predicted, through thermodynamic arguments, the possibility of forming giant vortices \cite{Christensson_NJP-2008,Yang_PRA2008}.


This large number of phenomena is due to the numerous experimental parameters that can be varied.  For example, the atom number, masses, interaction strengths, trapping frequency, and trap ellipticity can all be varied for each component separately.  The parameter space is far too large to be fully investigated through numerical simulations.  For this reason, we investigate the dynamics of a TCBEC through analytic methods then investigate points of interest through numerical simulations.  For single component BECs, considerable theoretical \cite{Recati_Stringari-PRL-2001,sinha-PRL-2001,Tsubota_PRA-2002,Lundh_PRA-2003,Kasamatsu_PRA-2003,Parker-PRA-2006,Corro-JPB-2007} and experimental \cite{Madison_PRL-2000,Madison_PRL-2001,Hodby_PRL-2001,Haljan_PRL-2001,Abo-Shaeer_Sci-2001} (for a summary see Ref.~\cite{Lundh_PRA-2003}) effort has been applied to understanding their dynamical properties under rotation.  A major result of this work was that only considering the thermodynamic stability of the BEC \footnote{The system is thermodynamically unstable if its free energy is not a global minimum.} does not correctly predict the onset of vortex nucleation.  It is instead necessary for the BEC to be dynamically unstable (dynamical instability implies thermodynamic instability, but not visa versa: see, for example, Ref.~\cite{Skryabin_PRA2000}).

We study the dynamical instabilities of TCBECs in rotating traps by deriving static solutions in the rotating frame, in the Thomas-Fermi limit.  We find that these solutions describe quadrupolar dynamics.  Through numerical simulations we show that instabilities in these solutions lead to phases that have already been predicted thermodynamically, such as interlocking vortex lattices.  In addition, these solutions predict interspecies-interaction-mediated centre-of-mass (COM) instabilities.  We observe these oscillations numerically and find that they can settle down into a stable symmetry broken state.  This state is different to previously-studied symmetry-broken states \cite{Esry_PRL-1997,Hall_PRL1998,Esry-Greene_PRA1999,Ohberg-Stenholm_PRA1998, Kasamatsu_Yatsui_Tsubota-PRA2001, Gordon-savage_PRA1998,Chui-Ao_PRA1999} in that it only occurs for attractive interspecies interactions.

The paper is laid out as follows.  In Sec.~\ref{sec:TFA} we derive equations that govern the stable irrotational (vortex free) motion of a TCBEC in a rotating trap.  We employ the Thomas-Fermi approximation to derive analytic results and find that when stable, both components undergo quadrupolar oscillations of different magnitudes.
In Sec.~\ref{sec:stability} we derive stability equations for these static solutions (critical points) in the rotating frame.  These instabilities indicate that the TCBEC will begin to evolve dynamically.  We identify four different instabilities:(i) catastrophic instability, (ii) ripple instability, (iii) COM instability, and (iv) intra-species COM instability. Types (i) and (ii) lead to turbulence and vortex nucleation.  Types (iii) and (iv) lead to COM motion.  In Sec.~\ref{sec:sims} we detail the results of numerical simulations used to investigate the analytic predictions.  These simulations reveal the dynamics induced by the instabilities studied in Sec.~\ref{sec:stability}.  They show how turbulence in the TCBEC allows for the formation of states with topologically distinct phase profiles which eventually settle down into either giant vortices, interlocking vortex lattices, or vortex sheets.  The simulations also show that the COM instabilities do not lead to vortex nucleation, but eventually settle down into a stable state (in the rotating frame), which, for the case of attractive interspecies interactions, breaks the $180^\circ$ rotational symmetry of the rotating frame Hamiltonian.

\section{Thomas-Fermi Approximation} \label{sec:TFA}

\subsection {The hydrodynamical equations}

In practice, a TCBEC can be stirred by introducing a rotating anisotropy into the confining potential.  If done adiabatically, the TCBEC will settle down into a state that oscillates in unison with the rotating trap.  The equations of motion can then be found by considering static solutions in the rotating frame.  A TCBEC can be described by 2 mean-field wavefunctions ($\Psi_1$ and $\Psi_2$) whose time evolution is dictated by the coupled two-component Gross-Pitaevskii Equation (GPE) \cite{Pethick-Smith}.  In a reference frame rotating with angular velocity $\bf \Omega$, these coupled equations become
\begin{eqnarray}
i \hslash\frac{\partial \Psi_j}{\partial t} \!\! &=& \!\! \Big[ \frac{\hslash^2 \nabla^2}{2 m_j} + {\bf V}_j + g_j |\Psi_j|^2  + g_{12}|\Psi_{j'}|^2 - {\bf \Omega} \cdot {\bf \hat{L}} \Big]\Psi_j \nonumber\\
\end{eqnarray}
where the $j$ subscripts ($j$ = $1$ or $2$) refer to the component under consideration, and $j'\ne j$.  $m_j$ and ${\bf V}_j$ are the mass and the potential affecting component $j$.  $g_j$ and $g_{12}$ are the intra and interspecies interaction coefficients given by  $g_j=4 n_0 \pi \hslash^2a_j/m_j$ and $g_{12}=2 n_0 \pi \hslash^2a_{12} [m_1+m_2]/[m_1 m_2]$.  Here, $a_j$ and $a_{12}$ are the intra and interspecies scattering lengths respectively.  The $n_0$ term allows for a rescaling such that $\parallel \Psi_j \parallel = N_j/n_0$ where $N_j$ is the number of atoms in component $j$.

As in the one-component case \cite{Recati_Stringari-PRL-2001}, it is possible to derive exact solutions in the Thomas-Fermi Approximation (TFA) (for a detailed description of the analytic methods, see \cite{Corro-JPB-2007}).  The GPE in the frame rotating with the potential is transformed using $\Psi_j =\sqrt{\rho_j({\bf r},t)}e^{i \phi_j({\bf r},t)}$ ($\rho_j$ is the density of component $j$, and  $\phi_j$ is the phase), and the TFA \cite{Baym_Pethick-PRL-1996} is applied, giving the hydrodynamical equations of motion:
\begin{eqnarray}
\frac{\partial \rho_j}{\partial t} &=& \nabla \cdot [ \rho_j(\frac{\hslash}{m_j}\gradphi- {\bf \Omega} \times {\bf
r}) ]\label{eq:hydro_rho}
\\
-\hslash \frac{\partial \phi_j}{\partial t} &=&\Theta_j +
g_{j}\rho_{j}+g_{12}\rho_{j'} \label{eq:hydro_mu}\\
\Theta_j & = & \frac{\hslash^2}{2 m_j} |\gradphi|^2 + V_j-\hslash \gradphi \cdot {\bf \Omega}
\times {\bf r} \label{eq:hydro_theta},
\end{eqnarray}

We take ${\bf \Omega}= \Omega(0,0,1)$ and  $V_j=\frac{m_j}{2} [ \left( 1 - \epsilon_j \right){\omperp_j }^2 x^2 +$$
\left( 1 + \epsilon_j \right) {\omperp_j }^2 y^2 + \omega_{z j}^2 z^2]$.  Here, $\epsilon_j$, $\omega_{zj}$ and $\omperp_j$ are respectively the ellipticity in the $x$-$y$ plane, the
frequency in the $z$ direction, and the frequency in the $x$-$y$ plane when $\epsilon_j=0$.  This gives a reference frame that is rotating around the z-axis in which the potential is static, which corresponds to modeling a potential that is rotating around the z-axis with angular velocity $-\Omega$. In what follows we enumerate the species so that $m_1^2 \omega_1^4/g_1 < m_2^2 \omega_2^4/g_2$.  We will see later that the behaviour of each component depends heavily on this criteria.

Steady state solutions in the rotating frame are obtained by setting $\frac{\partial \rho_j}{\partial t}$ =0 and $- \hslash \frac{\partial \phi_{j}}{\partial t} =
\mu_{j} \equiv$ the chemical potential of species $j$.  Solving Eq. (\ref{eq:hydro_mu}) gives two possible solutions for the density: if the density of both components is non zero, then
\begin{eqnarray}\label{eq:dens_2comp}
\rho_j^{o}= \frac{g_{j'} (\mu_j- \Theta_j)-g_{12}(\mu_{j'}
-\Theta_{j'})}{g_j g_{j'} - g_{12}}.
\end{eqnarray}
If one of the components has zero density, then the
 solution for the other component is
\begin{eqnarray}\label{eq:dens_1comp}
\rho_{j}^{s}= \frac{1}{g_j}(\mu_j- \Theta_j).
\end{eqnarray}

Regions in the BEC where Eqs.~(\ref{eq:dens_2comp}) and (\ref{eq:dens_1comp}) are applicable will be referred to as the \textit{overlapping region} and \textit{singular region} respectively (indicated by superscripts $o$ and $s$).  As in the one-component case, the TFA has the effect that $\rho_j$ can become negative \cite{Baym_Pethick-PRL-1996}.  When this is the case, we assume that  $\rho_j=0$.  Given the above, the total density for each component can be expressed as
\begin{eqnarray}
\rho_j=\rho_j^{o}H(\rho_j^{o})H(\rho_{j^{\prime}}^{o})+\rho_j^{s}H(-\rho_{j^{\prime}}^{o})H(\rho_{j}^{s}),
\label{eq:dens_total}
\end{eqnarray}
where $H(\rho)$=$0$ for $\rho<0$ and $H(\rho)$=$1$ for $\rho>0$.  Without rotation ($\Omega=\nabla \phi_1=\nabla \phi_2=0$), Eq.~(\ref{eq:dens_total}) corresponds to that derived in Refs.~\cite{Ho-PRL-1996, Riboli-PRA-2002}.

\subsection {Validity of the Thomas-Fermi Approximation} \label{sec:sep_phase_desc}

In this section we investigate the Thomas-Fermi approximation and its validity for a non-rotating TCBEC.  A two-component BEC has two phases.  When $-\sqrt{g_1 g_2}<g_{12}<\sqrt{g_1 g_2}$ the system is in the \textit{miscible phase} where both components interpenetrate. $g_{12}>\sqrt{g_1 g_2}$ is the \textit{immiscible phase}~\cite{Timmermans_PRL-1998,Svidzinksy_PRA2003} where the two components repel each other.  In the case of a harmonic trap, the interspecies repulsion causes either one component to form a shell around the other, or both components to separate asymmetrically \cite{Esry_PRL-1997} about the trap centre.  The asymmetric state occurs for $g_1 \approx g_2$, $m_1 \approx m_2$, and $g_{12}>\sqrt{g_1 g_2}$ \cite{Svidzinsky_PRA2_2003}.

\begin{figure}
\includegraphics[width=8.5cm,angle=0]{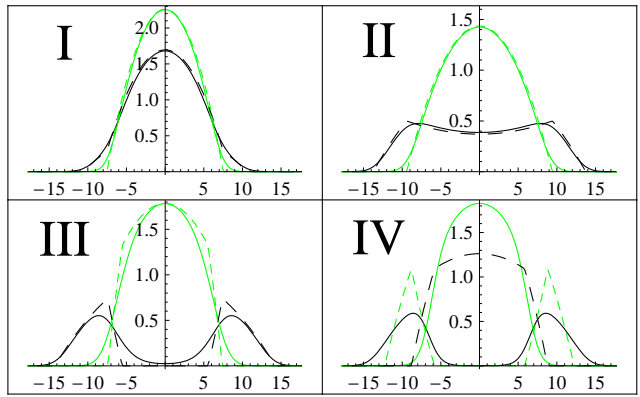}
\caption{\label{fig:densslices}(Color online) Examples of 1D slices of the TF density profiles (dashed curves) and exact numerical solutions of the two component GPE (solid curves) for component 1 (black), component 2 (green [grey]), $\epsilon_1=\epsilon_2=0$, $\omega_1=\omega_2=0.1$, $\hslash=1$, $m_1=1$, $m_2=1.5$, $g_1=1$, $g_2=0.5$. (I) $g_{12}=-0.5$, (II) $g_{12}=0.4$, (III) $g_{12}=0.65$,(IV) $g_{12}=0.8$.   The horizontal (vertical) axes are given in units of $l$ ($l^{-3}$), where $l=g_1 m_1/ \hbar^2$.}
\end{figure}

For BECs in a harmonic trap, the following general behaviour can be seen from Eq.~(\ref{eq:dens_2comp}).  When $g_{12}<g_2 m_1 \omega_1^2/(m_2 \omega_2^2)$ both components form overlapping density profiles with concave down parabolic shapes [Fig.~\ref{fig:densslices} (I)].  When $g_{12}=g_2 m_1 \omega_1^2/(m_2 \omega_2^2)$, $\rho_1^{o}$ is a constant, and, for values of $g_{12}$ larger than this, $\rho_1^{o}$ begins to dip in the overlapping region at the centre of the condensate [Fig.~\ref{fig:densslices} (II)].  As $g_{12}$ is increased, $\rho_1^{o}$ dips further until $g_{12}> g_2 \mu_2 /\mu_1$ where the TFA predicts $\rho_1^{o}=0$ at the very centre of the trap [Fig.~\ref{fig:densslices} (III)].

The Thomas-Fermi approximation assumes that the kinetic energy of the wavefunction is negligible compared to the interaction and potential energy terms.  For single-component BECs, it is valid for large g and atom number, which is satisfied for typical experimental parameters. These criteria must also be satisfied for each component individually if the TFA is applied to a TCBEC.  Moreover, it has been shown \cite{Pu_Bigelow-PRL-1998} that the accuracy of the TFA in the two-component case can have a more complicated dependence on the parameters used.  In this section, we explore the parameter space of a TCBEC to identify these additional constraints.

We have conducted extensive comparisons between the TFA and full GPE solutions, identifying the following conditions for TFA validity:  For $-\sqrt{g_1 g_2}<g_{12}<g_2 \mu_2 /\mu_1$ we find excellent agreement [Figs.~\ref{fig:densslices} (I) and (II)] ($g_{12}=g_2 \mu_2 /\mu_1$ is the point where the TFA predicts $\rho_1(\bm{r})$ reaches 0 at $\bm{r}=0$).  As $g_{12}=g_2 \mu_2 /\mu_1$ is approached the TFA results deviate from the full GPE results, but remain qualitatively correct for $g_2 \mu_2 /\mu_1<g_{12}<\sqrt{g_1 g_2}$ [Fig.~\ref{fig:densslices} (III)].  For $g_{12}>\sqrt{g_1 g_2}$ the TFA solutions are completely different from the full GPE solutions [Fig.~\ref{fig:densslices} (IV)].  This is due to the intense repulsion between the two components resulting in sharp curvature in the wavefunctions, and hence kinetic energy cannot be neglected.  For $g_{12}<-\sqrt{g_1 g_2}$ the TFA solutions are unphysical reflecting the fact that, without kinetic energy, the condensate is unstable.  In what follows we therefore consider the TFA solutions only in the region of their validity: $-\sqrt{g_1 g_2} < g_{12}< \sqrt{g_1 g_2}$.  In other words, the TFA is valid for any two-component BEC that is within the miscible phase. Eq.~(\ref{eq:dens_2comp}) shows that in this region of validity the component with the smallest value of $m^2 \omega^4/g$ (component 1 by definition) will always sit to the outside of component 2, and, as expected, this behaviour continues into the immiscible phase where component 1 forms an almost hollow shell around component 2.

One should note that the condensates begin to separate before the immiscible phase is reached, and the transition as $g_{12}$ passes $\sqrt{g_1 g_2}$ is continuous, i.e. there is no qualitative difference upon reaching the immiscible phase.  However, as shown by Timmermans \cite{Timmermans_PRL-1998}, the nature of the component separation is different in the two phases.  For $g_{12}<\sqrt{g_1 g_2}$ any separation is caused purely by the potential, if no trap gradient existed then both species would overlap with constant densities.  For $g_{12}>\sqrt{g_1 g_2}$ the two species separate irrespective of the trap.  Timmermans points out that this situation is much like the case for ordinary fluids.  Two fluids may mix freely but can still be separated by an external potential such as gravity.  They will, however, mix freely again once stirred.  Conversely, immiscible fluids such as oil and water always separate, and remain so even after mixing.  Indeed, we find that although there is no observable crossover between the two phases while the condensate is stable, once the BEC is stirred via rotation, the condensate displays very different behaviour depending on which phase it is in.

\subsection {Rotation}

In this section we derive the response of the condensate phase and density to rotation.  Without rotation, the wavefunction phase is constant for both components.  After introducing rotation, solutions to the equations of motion can be found by inserting a quadrupolar oscillation ansatz for the phase
\begin{equation}
    \phi_j=\frac{m}{\hbar}[\alpha_j^{o}H(\rho_j^{o})H(\rho_{j^{\prime}}^{o})+ \alpha_j^{s}H(-\rho_{j^{\prime}}^{o})H(\rho_{j}^{s}) ] x y,
\end{equation}
and Eq.~(\ref{eq:dens_total}) into Eq.~(\ref{eq:hydro_rho}) and solving for the $\alpha$'s.  For $\alpha_{j}^{s}$ this gives:
\begin{eqnarray}
{ {\alpha_{j}^{s}}}^3 - 2  {\alpha_{j}^{s}}
{\Omega }^2 + { {\omperp_{j}}}^2 \left(  {\alpha_{j}^{s}} -
{\epsilon_{j}} \Omega \right)=0,&& \label{eq:alpha1pol}
\end{eqnarray}
which has up to 3 real solutions and is identical to the one component case \cite{sinha-PRL-2001}.  For the $\alpha^{o}$'s we get two simultaneous equations:
\begin{eqnarray}
- 2  {g_{12}} {m_j} \left( { {\omperp_j}}^2 \left(  {\alpha_{j'}^{o}} -
{\epsilon_j} \Omega \right) +  {\alpha_j^{o}} \left(  {\alpha_j^{o}}
{\alpha_{j'}^{o}} - 2 {\Omega }^2 \right)  \right) &&\nonumber\\
+ 2 {g_j} {m_{j'}} \left( { {\alpha_{j'}^{o}}}^3 - 2  {\alpha_{j'}^{o}}
{\Omega }^2 + { {\omperp_{j'}}}^2 \left(  {\alpha_{j'}^{o}} -
{\epsilon_{j'}} \Omega \right) \right)=0.&& \label{eq:alpha2pol}
\end{eqnarray}
This yields up to nine real solutions for $\{\alpha_1^{o},\alpha_2^{o}\}$.

The $\alpha$'s are real constants representing the magnitude and orientation (positive or negative) of quadrupolar oscillation in the two components.  Specifically, the singular and overlapping regions of component 1 (component 2) undergo quadrupolar oscillations of different magnitudes, labeled $\alpha_{1}^{s}$ and $\alpha_{1}^{o}$ ($\alpha_{2}^{s}$ and $\alpha_{2}^{o}$) respectively.  The density profile of each component is heavily influenced by this oscillation, in particular its ellipticity.  This dependence has the interesting property that when $\alpha$ is negative, the  BEC is deformed oppositely to the elliptical deformation caused by the trap.  In other words the BEC forms an elliptical profile that is rotated $90^\circ$ to that of the confining elliptical potential.

\begin{figure}
\includegraphics[width=8.5cm,angle=0]{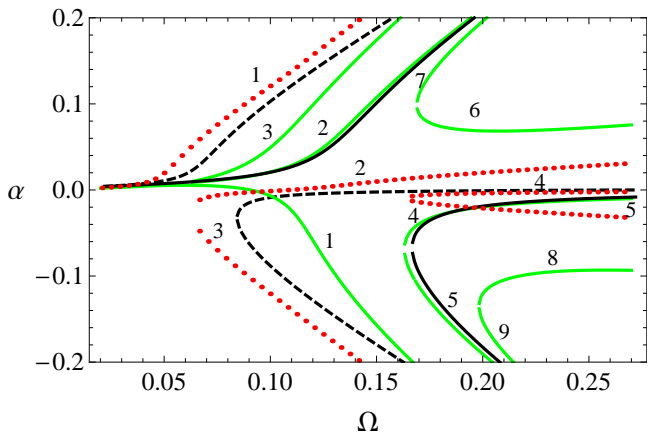}
\caption{\label{fig:alphaplots}(Color online) The solutions for the $\alpha$'s with $\hslash=\mu_1=\mu_2=1, \omega_1=0.2, \omega_2=0.1, \epsilon_1=\epsilon_2=.1, m_1=1, m_2=1.5, g_1=1, g_2=0.5, g_{12}=0.1$.  Shown are solutions to $\alpha_1^{o}$ (thick green [thick grey]), $\alpha_2^{o}$ (red dots [grey dots]), $\alpha_1^{s}$ (black dashed), $\alpha_2^{s}$ (black line).  The branch numbering sits above the corresponding branch, it allows matching of each $\alpha_1^{o}$ solutions to the corresponding $\alpha_2^{o}$ solution.  $\Omega$ and $\alpha$ are in units of $\mu_1/\hbar$.}
\end{figure}

\subsection{Discussion}
An example with 7 solutions to Eq.~(\ref{eq:alpha2pol}) for $\{\alpha_1^{o},\alpha_2^{o} \}$ is shown in Fig.~\ref{fig:alphaplots}. Each branch represents a different static solution for a rotating TCBEC, each with quadrupolar oscillations of different magnitudes and orientations.  Although these static solutions display a very complicated dependence on the many free parameters in the system, the situation is greatly simplified by the fact that the BEC will prefer the branch with the lowest energy.

The energy of a given solution increases as the magnitude of $\alpha$ increases.  Also,
for the solutions given by Eq. (\ref{eq:dens_2comp}), the density of component $i$ in the overlapping and singular regions does not join continuously at the boundary unless $\alpha_i^{o}=\alpha_i^{s}$.  This adds a large amount of kinetic energy at the interface of the two regions as the density profile must vary rapidly to join the two regions.  The solution with the smallest energy can therefore be determined by removing $\alpha_i^{o}$ branches that do not correspond closely to one of the $\alpha_i^{s}$ branches.  One then selects the remaining branch which has the smallest magnitude.  From the figure, it is evident that for $\Omega<0.06$ branch 1 has the lowest energy, for $0.06<\Omega<0.16$ branch 2 has the lowest energy, and for $\Omega>0.16$ branch 4 has the lowest energy.

The behaviour of the BEC can then be described as follows.  Initially, after condensation and before introducing rotation, the BEC will be in a state that corresponds to the stable static solution which has the lowest energy for the given set of experimental parameters.  The BEC can then be transferred to a new state by adiabatically ramping these parameters.  As this is done, the BEC will move along one of the static solutions. In the case of Fig.~\ref{fig:alphaplots}, each of branches 1, 2, and 4 can be accessed by ramping the parameters in different ways.  Branch 1 can be accessed by keeping $\epsilon$ constant while ramping $\Omega$ from $0$ to some final value or by keeping $\Omega<0.06$ fixed and ramping $\epsilon$.  Branch 2 (4) can be accessed by fixing $0.06 < \Omega < 0.16$ ($\Omega>0.16$) and ramping $\epsilon$ from $0$ to some final value.  Branch 2 is also of particular interest because it leads to each BEC component undergoing quadrupolar oscillations in opposite directions (i.e. $\alpha_1^{o}>0$ and $\alpha_2^{o}<0$).

The BEC cannot follow these static solutions for all parameter values; eventually, one of two possibilities occurs.  In the first case, the static solution which the BEC is following can cease to exist (\emph{catastrophic instability}) \cite{Parker-PRA-2006}.  This leads to a massive disruption in the BEC's density profile, followed by the onset of turbulence and vortex nucleation.  The second case occurs when the BEC is still in a state described by a static solution, but the solution itself is not stable.  In this case, the BEC can either become turbulent (\emph{ripple instability}) which leads to the formation of vortices, or the COM of the BEC becomes unstable (\emph{centre of mass instability}).

\section{Stability}\label{sec:stability}
\subsection{Equations}\label{sec:insteq}
Stability can be analysed by linearising Eqs. (\ref{eq:hydro_rho},\ref{eq:hydro_mu}) about
the critical points.  We consider infinitesimal perturbations $\rho_j \rightarrow
{\rho_0}_j+\delta \rho_j$ and $\phi_j \rightarrow {\phi_0}_j+\delta
\phi_j$, with \{${\rho_0}_i$, ${\phi_0}_i$\} being a set of static solutions to Eqs. (\ref{eq:hydro_rho},\ref{eq:hydro_mu}).  In the overlapping region where both condensates coexist, one obtains
\begin{eqnarray}
\frac{\partial}{\partial t} \left[
\begin{array}{c}
\delta {\phi}_1 \\
\delta {\rho}_1 \\
\delta {\phi}_2 \\
\delta {\rho}_2 \\
\end{array}
\right]
& = &\left[ \begin{array}{c c}
\bm{A_1} & \begin{array}{c c} 0 & \frac{g_{12}}{\hslash}  \\ 0 & 0 \end{array} \\
\begin{array}{c c} 0 & \frac{g_{12}}{\hslash}  \\ 0 & 0 \end{array} & \bm{A_2}
\end{array} \right]
 \left[
\begin{array}{c}
\delta {\phi}_1 \\
\delta {\rho}_1 \\
\delta {\phi}_2 \\
\delta {\rho}_2 \\
\end{array}
\right] \label{eq:insteq}
\\
\bm{A_j} & = & \left[ \begin{array}{c c}
{\bf v}_j \cdot \nabla & \frac{g_j}{\hslash}\\
\nabla \cdot ({\rho_0}_j \frac{\hslash}{m_j} \nabla) &{\bf v}_j
\cdot \nabla
\end{array} \right], \label{eq:insteqA}
\end{eqnarray}
and in the singular region one obtains
\begin{eqnarray}
\frac{\partial}{\partial t} \left[
\begin{array}{c}
\delta {\phi}_j \\
\delta {\rho}_j \\
\end{array}
\right]
& = & \bm{A_j}
 \left[
\begin{array}{c}
\delta {\phi}_j \\
\delta {\rho}_j \\
\end{array}
\right] \label{eq:insteqouter}.
\end{eqnarray}
Here ${\bf v}_j = \frac{\hslash}{m_j} \nabla {\phi_0}_j -{\bf
\Omega} \times {\bf r}$ is the wave function velocity in the rotating frame at position ${\bf r}$.  ${\rho_0}_j$ is given by Eq.~(\ref{eq:dens_2comp}) in the overlapping region and Eq.~(\ref{eq:dens_1comp}) in the singular region.

As in the one component case \cite{sinha-PRL-2001}, we find
that the eigenfunctions of the collective mode equations [Eqs.~(\ref{eq:insteq} - \ref{eq:insteqouter})] are polynomials of the form $\delta \rho_j = \sum_{pqr}  \beta_{j p q r}
x^{p}y^{q}z^{r}$, $\delta \phi_j =\sum_{pqr} \gamma_{j p q r}
x^{p}y^{q}z^{r}$, where $\beta_{j p q r}$ and $\gamma_{j p q r}$ are constants. The BEC is
unstable when one of the eigenvalues has a positive real part, meaning that small perturbations about the static solutions grow exponentially.

As well as the above method, extra information can be obtained by investigating the stability of the overlapping region of each component separately.  This is done by substituting  Eq.~(\ref{eq:dens_2comp}) [instead of Eq.~(\ref{eq:dens_1comp})] into Eq.~(\ref{eq:insteqouter}).  By doing so, one can see in which component and in which region an instability originates.  This, however, neglects interspecies cross terms; it is equivalent to treating the interaction between components as a static potential.  Interestingly, we find that this second method gives more accurate results than if cross terms are considered.  We take this to be an indication that the collective mode cross interactions are not well described within the TFA framework.

\subsection{Different types of instabilities}
Calculating the eigenvalues to Eq.~(\ref{eq:insteq}) gives the regions of instabilitiy in the parameter space. Valuable information regarding the nature of the instability can also be gained by observing the corresponding eigenvectors.  This allows the instabilities of a rotating two-component BEC to be divided into four different types based on the very different effects they have on the overall structure of the BEC.  Simulations showing the effects of these instabilties in detail are given in the next section.
\subsubsection{Classical COM instability}
In its original static configuration, the density of BEC component $j$ is a quadratic of the form (for simplicity of the argument we ignore trap anisotropy and the distinction between the overlapping and singular region) $\rho_j = \mu_j-x^2_j-y^2_j-z^2_j$.  Consider the case where an eigenvalue $\lambda$ corresponds to an eigenvector first order in a position coordinate, e.g. one of the form $\delta \rho_j = \Delta_x \,x_j$.  This perturbation will have the effect of displacing the centre of mass of the BEC along the $x$-axis, i.e. $\rho+\delta \rho = (x_j+.5\Delta_x)^2+y^2_j+z^2_j + \mu'_j$.  If $\lambda$ has a positive real part then this perturbation will grow in magnitude, displacing the centre of mass of the BEC even further and hence causing a \emph{COM} \emph{instability}.  We find that component $1$ always has such an instability in the range $\omperp_1\sqrt{1-\epsilon_1}< \Omega<\omperp_1\sqrt{1+\epsilon_1}$ regardless of the other parameters.  This is in fact the same instability as experienced by a classical point particle in a rotating harmonic trap.  It is caused by the rotation frequency coupling to the oscillation frequency and is also experienced by a one-component BEC \cite{Recati_Stringari-PRL-2001}.  In this case, it either causes the BEC to oscillate as a whole about the trap centre, or it can drive the BEC out of the trap \cite{Corro-JPB-2007}.  In a two-component BEC, it can still occur provided there is little interaction between the components.  This is discussed in more detail in Sec. \ref{sec:TFA_results}.
\subsubsection{Intra-species COM instability}
The above classical COM instability affects each component separately and independently of the interspecies interactions.  We find another class of COM instability that occurs due to the interaction of the two components and can have a profoundly different effect on the BEC.  They are again predicted by the instability of a perturbation first order in a position coordinate, but are differentiated from the first type in that they appear in the overlapping region of the condensate where both components are interacting.  These \emph{intra-species COM instabilities} are due purely to interactions of the superfluid components and lead to instability in the COM of each component separately, but the total COM of the condensate remains stable.  They result in interesting dynamics which are described in section \ref{sec:sims}
\subsubsection{Ripple Instability}
Perturbations of quadratic or higher in the position coordinates represent ripples through the phase and density profile of the BEC.  If these perturbations are unstable they directly disrupt the smooth quadratic profile.  These \emph{ripple instabilities} lead to turbulence and vortex nucleation.
\subsubsection{Catastrophic Instability}
In this case, the static solution that the BEC was following during adiabatic ramping ceases to exist.  Perturbations of all orders are unstable and the BEC is torn apart in a spectacular fashion \cite{Parker-PRA-2006}.  After the initial onset, the BEC becomes turbulent, and, as in the ripple instability, vortices nucleate.

\subsection{Results} \label{sec:TFA_results}
\begin{figure}
\includegraphics[width=8cm,angle=0]{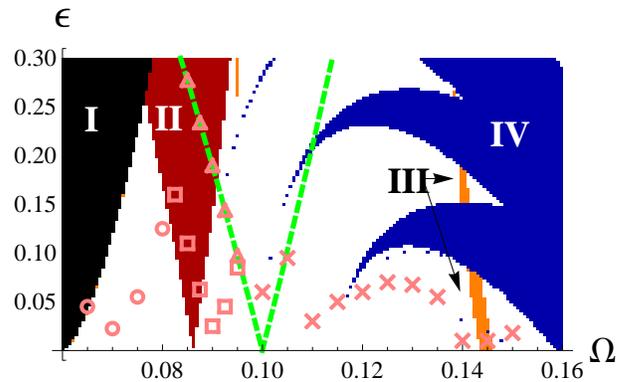}
\vspace{-0.0cm}
\caption{\label{fig:sp_instab}(Color online) The instability regions within the parameter space for a system with $\hslash=\mu_1=\mu_2=1, \omega_1=0.2, m_1=1, g_1=1, \omega_2=0.1, m_2=1.5, g_2=0.5, g_{12}=.1$, catastrophic instability (Region I - black), intra-species COM instability (Region II - red [dark grey]), ripple instability in component 2 (Region III - orange [light grey]), ripple instability in component 1 (Region IV - blue [dark grey]), classical COM instability in component 2 (green [light grey] dashed line).  Marks show the results for instabilities found using GPE simulations for catastrophic instabilities (circles), intra-species COM instabilities (squares), classical COM instabilities (triangles), ripple instabilities (crosses). $\Omega$ is in units of $\mu_1/\hbar$.
}
\end{figure}

\begin{figure}
\includegraphics[width=8cm,angle=0]{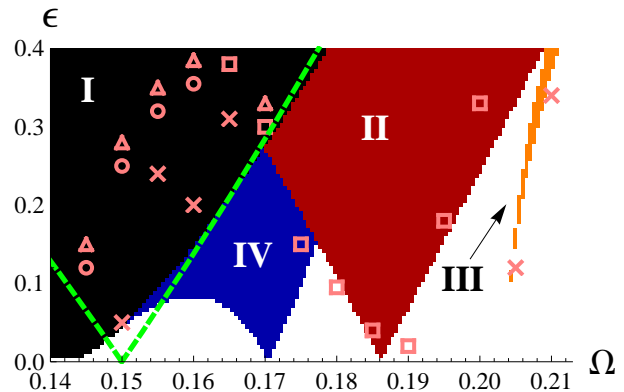}
\vspace{-0.0cm}
\caption{\label{fig:at_instab}(Color online) same as Fig.~\ref{fig:sp_instab} but with $\hslash=\mu_1=\mu_2=1, \omega_1=0.1, m_1=1, g_1=0.5, \omega_2=0.15, m_2=1, g_2=1, g_{12}=-.3$.
}
\vspace{-0.0cm}
\end{figure}

Using Eq.~(\ref{eq:insteq}) we present two examples (Figs. \ref{fig:sp_instab} and \ref{fig:at_instab}) of the instabilities predicted by the TFA and how they appear in different regions of the parameter space.  We have used the second method described in Sec.~\ref{sec:stability}.  On the same figures are plotted the point where the BEC becomes unstable in GPE simulations conducted for the same set of parameters.  In general the solutions and their stability can be evaluated for any ramping procedure, the only difference being that different ramping paths can accesses different $\alpha$ branches.  In our GPE simulations we fix $\Omega$ and then ramp $\epsilon_1=\epsilon_2$ from $0$.  This allows us to investigate regions of the phase diagram that, for example, would not be accessible by ramping $\Omega$ for fixed $\epsilon$.  The point of instability is then determined from the simulations using the method presented in \cite{Corro-JPB-2007}.

The results of the simulations show how the TFA results can be used to predict BEC behaviour.  The methods derived above successfully pick up the different regions in the parameter space where different instabilities occur.  Sometimes the results of the GPE simulations are exactly predicted in the TFA, at other times the phenomenology is as predicted but its location numerically shifted in $\Omega$ by as much as $15\%$ .  As discussed above, the TFA is valid within the bulk of the condensate.  The shifting is an indication that for the chosen parameters, the boundary of the BEC is affecting the particular instability that has been shifted.

For the case of repulsive interspecies interaction ($g_{12}>0$), we find that the stability of the overlapping region of component 1 is highly dependent on its connection to the singular region [see, for example, the non rotating case Fig.~\ref{fig:densslices}(III)].  The results using the TFA on the overlapping region of component 1 are only accurate when $g_{12}<0$ and should not be used for $g_{12}>0$.

Fig.~\ref{fig:sp_instab} shows the instabilities found on branch 2 of Fig. \ref{fig:alphaplots}.  A BEC on this branch can, depending on the parameters, experiences either a intra-species COM, ripple, catastrophic, or classical COM instability.   The actual position of the catastrophic instability is shifted to higher $\Omega$ from the TFA prediction by $\sim 10\%$ and the intra-species COM instability is shifted to higher $\Omega$ by $~5\%$.  The green dashed line indicates the region where the classical COM instability would be for component 2 were the other component not present.

Fig.~\ref{fig:at_instab} focuses on the COM instability for a different system.  This system has attractive interspecies interactions and equal masses but different trapping frequencies, which could correspond, for example, to two condensates of the same atomic species but different hyperfine states.  The intra-species COM instability and ripple instability in component 2 are exactly as predicted.  The catastrophic instability is shifted to lower $\Omega$ from the predictions by $~10\%$.  The ripple instability in the singular region of component 1 is similar to, but much less influential than, the prediction.  However this is not surprising because for attractive interspecies interactions the two components are pulled tightly together and the singular region is almost non-existent [see Fig.~\ref{fig:densslices} (I)].

The TFA predictions for both examples show that the interspecies interactions should halt the onset of the classical COM instabilities when both species are performing stable quadrupolar oscillation.  Interestingly, the intra-species COM instability pulls the two components apart leaving them once again susceptible to the classical COM instability.  If this has occurred before reaching, or whilst within the classical COM instability, the component in question will exit the trap as though the other component were not present.  Likewise, the catastrophic instability causes both components to be wrenched apart. If one of the components is within a classical COM instability during the break down, it is free of the other component for long enough that it will exit the trap.  This effect is clearly seen on Figs~\ref{fig:sp_instab} and \ref{fig:at_instab}.

\section{Simulations} \label{sec:sims}

To compliment these Thomas-Fermi results we have adapted 2D Truncated Wigner simulations \cite{Steel-PRA-1998} to the two-component case, and investigated the instabilities predicted.  Interesting results are attained for each different type of instability, and these can be understood in terms of the TFA results.  Also, the numerical methods allow us to extend the investigation to the immiscible phase where the TFA is no longer valid.

As well as the results shown in Figs.~\ref{fig:sp_instab} and \ref{fig:at_instab} we conduct simulations for parameters that correspond to an $^{87}$Rb-$^{133}$Cs system \cite{Harris-JPB-2008} and an $^{87}$Rb-$^{85}$Rb system \cite{Papp_PRL-2008}.  These systems are of particular interest because the scattering lengths of $^{133}$Cs and $^{85}$Rb can be tuned via a Feshbach resonance \cite{Inouye_Nature-1998,Papp_PRL-2008}.  This means that both the immiscible and miscible phases can be accessed experimentally.  The $^{87}$Rb-$^{85}$Rb is also important because it satisfies the requirements for the creation of vortex sheets.  The $^{87}$Rb-$^{133}$Cs system is used to simulate the creation of the other rotating states.  For the $^{87}$Rb-$^{133}$Cs system, $a_{12}$ is unknown.  We, therefore, give examples of simulations conducted for a number of different choices of $a_{12}$.

\subsubsection{The truncated Wigner method}

The truncated Wigner method simulates quantum vacuum fluctuations by adding appropriate classical random fluctuations to the coherent field of the BEC's initial state. In this system the fluctuations serve two purposes. Firstly, the fluctuations provide a seed of noise to break the BEC symmetry when the quadrupolar oscillation becomes unstable. Secondly, they enable incoherent scattering processes to occur, by which condensate atoms are scattered into a thermal cloud. This method has been used to describe, for example, the formation of scattering halos in condensate collisions \cite{norrie-PRL2005,scott-PRA2006}, and the suppression of Cherenkov radiation \cite{scott-PRA2008}. Hence the turbulent BEC can relax into a rotating eigenstate, such a vortex lattice, in contrast to the bare GPE which conserves energy and atom number.

In practice, the fluctuations are included as follows. The initial wavefunction for each species is obtained by solving the time-independent two-component GPE. These two wavefunctions are then expanded over a plane-wave basis, with a maximum cutoff wavevector to prevent Fourier aliasing. Quantum fluctuations are introduced into each wavefunction separately by adding random complex noise to each plane-wave mode. The amplitude of the quantum fluctuations has a Gaussian distribution, with an average value of half a particle \cite{norrie-PRL2005}.  For a thorough description of the method, see Refs.~\cite{Steel-PRA-1998,scott-PRA2006}.

\subsection{Results}

\begin{figure}
\includegraphics[width=8.5cm,angle=0]{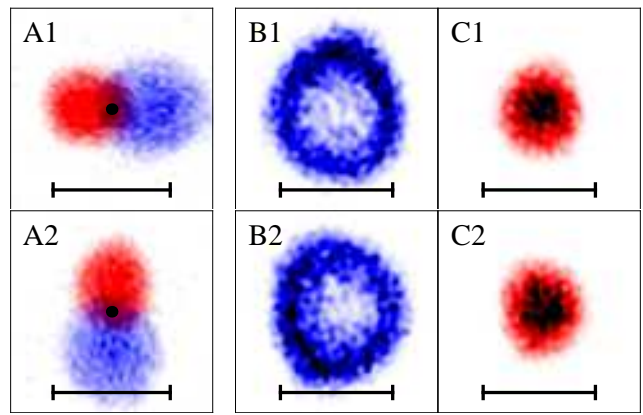}
\vspace{-0.0cm}
\caption{\label{fig:simscomi}(Color online) Plots of the density for BECs undergoing intra-species COM instabilities for an $^{87}$Rb (component 1 - blue [panels B1 B2, right half of A1 and bottom half of A2]) - $^{133}$Cs (component 2 - red [panels C1 C2, left half of A1 and top half of A2]) system with $a_{1}=a_{2}=5.4$ nm, $\omega_1=\omega_2=3.15$ Hz.  Attractive case ($g_{12}<0$): $a_{12}=-0.65$ nm, $\Omega=3.66$ Hz.  (A2) is the density shortly after (A1) showing the two components rotating around each other.  The black dot is the COM of the system.  Repulsive case ($g_{12}>0$): $a_{12}=3.9$ nm, $\Omega=3.66$ Hz.  (B1, C1) is the density shortly after (B2,C2) showing component 2 (C1,C2) bouncing of the shell of the enclosing component 1 (B1,B2).  The horizontal bars denote 40 $\mu$m.}
\vspace{-0.0cm}
\end{figure}

\begin{figure*}[ht!]
\includegraphics[width=.95\textwidth]{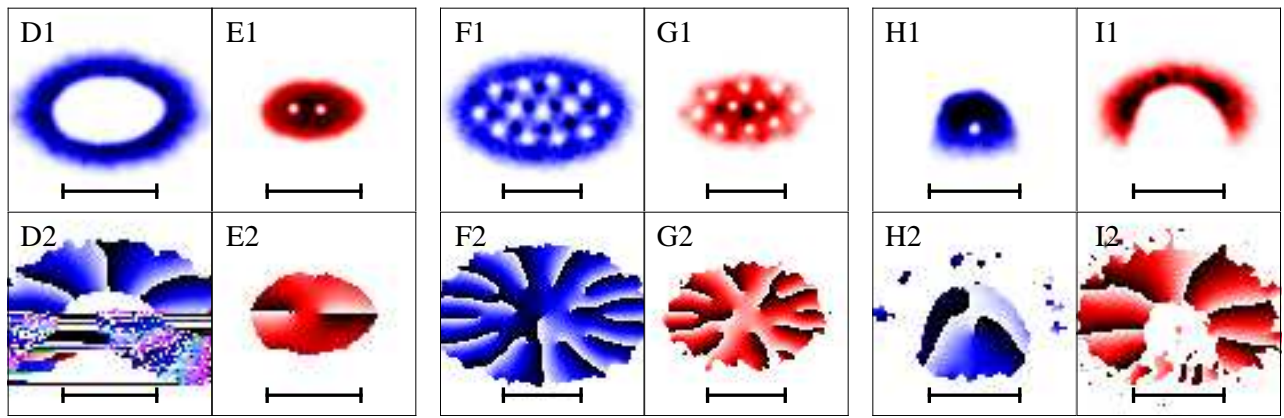}
\caption{(Color online) Results of 2D Truncated Wigner simulations ending in vortex nucleation. Panels D,E,F,and G show an $^{87}$Rb (component 1) - $^{133}$Cs (component 2) system with $a_{1}=a_{2}=5.4$ nm, $\omega_1=\omega_2=3.15$ Hz.  (D,E) $a_{12}=7$ nm, (F,G) $a_{12}=3.3$ nm.  Panels H and I show an $^{87}$Rb (component 1) - $^{85}$Rb (component 2) system with $a_{1}=5.24$ nm, $a_{2}=11.27$ nm, $a_{12}=11.3$ nm, $\omega_1=\omega_2=3.15$ Hz.  The horizontal bars denote 40 $\mu$m.  The number of atoms in each simulation is such that the peak density of $^{87}$Rb is $4 \times 10^{18}$ m$^{-3}$ and both components have equal norms.  D1,F1, and H1 (E1,G1, and I1) is the density and D2,F2, and H2 (E2,G2, and I2) the phase of component 1 (2).  The simulations are conducted by ramping $\epsilon$ while fixing $\Omega$ at 1.75 Hz}\label{fig:simsvort}
\vspace{-0.6cm}
\end{figure*}

\subsubsection{Intra-species COM instability}
The TFA results show that as $g_{12}$ is increased or decreased from $0$, the classical COM instability of component 2 begins to shift and becomes an intra-species COM instability.  For $g_{12}<0$ the simulations show that when this instability is reached, the COM of the entire BEC is stable; however, the COM of the individual components becomes unstable and the two components separate.  The instability is caused purely by interspecies interactions, so as the components separate and interactions decrease, the trap pushes the two components back together again.  The effect is that both components begin oscillating; eventually, the BEC settles down into a stable state with both components orbiting around one another [Fig.~\ref{fig:simscomi} (A)].  This state is static in the rotating frame and implies a breaking of the $180^\circ$ symmetry of the rotating frame Hamiltonian.

For $g_{12}>0$ component 2 sits within component 1.  When the intra-species COM instability in component 2 is reached, it is still trapped within component 1; it begins bouncing off the surrounding shell [Fig.~\ref{fig:simscomi} (B,C)].  Eventually, the two components disrupt each other and settle down into a state that has component 2 still sitting within component 1, but with both components significantly diffused into one another.  During these instabilities, no vortices are formed.

\subsubsection{Classical COM instability}
The classical COM instability can affect a TCBEC if both components experience it at the same time (i.e. if $\omega_1 \approx \omega_2$).  However, if only one component is in a regime of classical COM instability then the TFA results predict, and simulations confirm, that this instability is suppressed during stable rotational motion.  None-the-less, as already discussed in Sec.~\ref{sec:TFA_results}, each component can still experience its classical COM instability independently of the other component provided it is freed from the other component first via a catastrophic or intra-species COM instability.

Another case where a classical COM instability can occur is near or in the immiscible phase where both components have a large singular region.  In this case, component 2 forms a tightly packed ball within the shell of component 1.  When component 2 reaches its classical COM instability it builds up enough momentum to break through component 1 while component 1 remains stable.  Component 2 then either exits the trap or begins to oscillate within the trapping potential.  The result is similar to a wrecking ball as the highly dense component 2 smashes through the dilute component 1.
\subsubsection{Catastrophic and ripple instabilities}
In the miscible phase, the simulations show that when a ripple or catastrophic instability is reached the BEC becomes turbulent and vortices enter.  They then form an interlaced vortex lattice [Fig.~\ref{fig:simsvort} (F,G)].  This state is of the same type as has been seen to form in experiments when a one-component BEC with a vortex lattice already present was split into two hyperfine components \cite{Schweikhard-PRL-2004}.

We also investigated the behaviour of the rotational instabilities near to and within the immiscible phase.  When $g_{12}$ is close to but less than $\sqrt{g_1 g_2}$, component 1 forms an almost hollow shell around component 2 even though the immiscible phase has not been reached.  As expected from the discussion at the end of Sec.~\ref{sec:sep_phase_desc}, once the condensate is stirred and turbulence reached, the two components mix together and a vortex lattice is formed.

Once within the immiscible phase however, there is a discontinuous change in the behaviour of a rotated BEC, and, importantly, the two components do not mix.  We find that once turbulence is induced, two different possible configurations with non trivial phase topology form.  The first is a vortex sheet which has been shown to be preferable thermodynamically in regions of parameter space where $g_1 \approx g_2$, $m_1 \approx m_2$, and $g_{12}>\sqrt{g_1 g_2}$ \cite{Kasamatsu_PRL2003,Woo_PRAR2007} [Fig.~\ref{fig:simsvort} (H,I)].  In this configuration the two components form separate domains of high density.   We find that the parameters required for the dynamical formation of a vortex sheet matches that of the thermodynamic analysis provided an instability has been reached; without an instability they do not form.  These conditions for vortex sheet formation also coincide with the parameters required for a symmetry broken initial state in the absence of rotation \cite{Svidzinsky_PRA2_2003}.

The second state we find is a giant vortex which has also been predicted thermodynamically \cite{Christensson_NJP-2008,Yang_PRA2008}.  We find that this state forms for any set of parameters that fall within the immiscible phase but do not match the requirements for the formation of vortex sheets.  As with the vortex sheets, a ripple or catastrophic instability must first be induced in order for this state to form dynamically.  The simulations show that once instability is reached multiple vortices push their way through the outer shell of component one and into the low density region in the middle of the trap where they congregate forming a giant vortex [Fig.~\ref{fig:simsvort} (D,E)].  Normally, two overlapping vortices are predicted to be thermodynamically unstable \cite{Castin_Dum-EUPD-1999}.  However, in this system the large density of component 2 attracts them to the centre while the outer shell of Rb holds them in.  These results show how in practice such a state could be created and that this state is a product of the immiscibility of the two components.

In summary, the above results show that there is a direct connection between the state of the non-rotating TCBEC and the state the rotating TCBEC will settle down into after instability has been reached.  That is, a miscible phase leads to vortex lattice after rotating, a symmetry broken state leads to vortex sheets, and an immiscible phase (other than the symmetry broken phase) leads to a giant vortex.

\section{Conclusion}
These results illustrate that BEC mixtures produce a rich variety of dynamical regimes that may be accessed by tuning experimental parameters.  We have examined the applicability of the TFA to a TCBEC and extended it to the rotating case allowing us to find symmetry breaking COM oscillations that are induced by interspecies interactions. In addition, the results give conditions under which interlaced vortex lattices, giant vortices and vortex sheets will spontaneously form.  The method allows all these phenomena to be understood and classified through particular instabilities occurring in different parts of the total condensate.

\begin{acknowledgments}
We thank Simon Cornish for helpful discussions and acknowledge funding from the ARC (IC and AMM) and the EPSRC (RGS).
\end{acknowledgments}

\bibliography{twincomprot}
\end{document}